\documentclass[usenatbib]{mn2e}

\newcommand{\hoz}{\rm H_2 ~ v = 1 \rightarrow 0 ~ S(1)}
\newcommand{\hto}{\rm H_2 ~ v = 2 \rightarrow 1 ~ S(1)}
\newcommand{\hozs}{\rm v = 1 \rightarrow 0 ~ S(1)}
\newcommand{\htos}{\rm v = 2 \rightarrow 1 ~ S(1)}
\newcommand{\kms}{\rm km\;s^{-1}}
\newcommand{\cmv}{\rm cm^{-3}}
\newcommand{\intensity}{\rm \times\;10^{-18}\;W\;m^{-2}\;arcsec^{-2}}

\title[Interaction between Sgr A East and GMCs]
{Interaction between the Northeastern Boundary of Sgr A East and Giant Molecular Clouds}

\author[Lee et al.]
{Sungho Lee$^{1}$,
 Soojong Pak$^{1,2}$\thanks{E-mail: soojong@astro.snu.ac.kr},
 Christopher J. Davis$^{3}$,
 Robeson M. Herrnstein$^{4}$,
 \newauthor
 T. R. Geballe$^{5}$,
 Paul T. P. Ho$^{4}$,
 and
 J. Craig Wheeler$^{6}$\\
 $^{1}$Astronomy Program in SEES, Seoul National University,
      Shillim-Dong, Kwanak-Gu, Seoul 151-742, South Korea\\
 $^{2}$Korea Astronomy Observatory,
      Whaam-Dong, Youseong-Gu, Taejeon 305-348, South Korea\\
 $^{3}$Joint Astronomy Centre, University Park,
      660 North A'ohoku Place, Hilo, HI 96720, USA\\
 $^{4}$Harvard-Smithsonian Center for Astrophysics,
      60 Garden Street, Cambridge, MA 02138, USA\\
 $^{5}$Gemini Observatory, 670 N. A'ohoku Place, Hilo, HI 96720, USA\\
 $^{6}$Astronomy Department, University of Texas, Austin, TX 78712, USA\\
}

\begin{document}

\date{Accepted 2002 ?? ??. Received 2002 ?? ??; in original form 2002 ?? ??}

\pagerange{\pageref{firstpage}--\pageref{lastpage}} \pubyear{2002}

\maketitle

\label{firstpage}

\begin{abstract} We have detected the $\hozs ~ (\lambda = 2.1218 \micron)$
and $\htos ~ (\lambda = 2.2477 \micron)$ lines of H$_2$ in the Galactic
centre, in a $90 \times 27$ arcsec region between the northeastern
boundary of the non-thermal source, Sgr A East, and the giant molecular
cloud (GMC) M-0.02-0.07. The detected $\hoz$ emission has an intensity of
1.6 -- 21 $\intensity$ and is present over most of the region. Along with
the high intensity, the broad line widths (FWHM = 40 -- 70 $\kms$) and
the $\hto$ to $\hozs$ line ratios (0.3 -- 0.5) can be best explained by a
combination of C-type shocks and fluorescence.  The detection of shocked
H$_2$ is clear evidence that Sgr A East is driving material into the
surrounding adjacent cool molecular gas. The H$_2$ emission lines have two
velocity components at $\sim$ +50 $\kms$ and $\sim$ 0 $\kms$, which are
also present in the NH$_3$(3,3) emission mapped by \citet*{mcg01}.  This
two-velocity structure can be explained if Sgr A East is driving C-type
shocks into both the GMC M-0.02-0.07 and the northern ridge of
\citet*{mcg01}.

\end{abstract}


\section{Introduction}

Sgr A East has frequently been interpreted as a supernova remnant due to
its shell structure and non-thermal spectrum (\citealt{jon74};
\citealt{gos83}
and references therein; and see the more recent references in
\citealt{mae02}). Some recent research, however, has
suggested that the energetics, size, and elongated morphology ($3 \times
4$ arcmin or $7 \times 9$ pc at $d = 8.5$ kpc) of Sgr A East cannot have
been produced by a typical supernova \citep{yus87,mez89}. \citet{mez89}
estimate the required energy to produce Sgr A East to be more than $4
\times 10^{52}$ ergs.  Modeling of the entire spectrum of Sgr A East by
\citet{fat99}, which fits very well with the observations of the
non-thermal emission of Sgr A East and EGRET $\gamma$-ray sources,
supports the energy estimate by \citet{mez89}.  Those authors concluded
that a single supernova explosion could explain the existence of Sgr A
East only if it occurred within the cavity formed by the stellar wind from
a progenitor star. In that scenario, however, the formation of the cavity
takes too much time ($\sim 10^6$ yr) compared with the orbital period
($\sim 10^5$ yr) of matter circling around the Galactic centre
\citep{mez89}.

\begin{figure*}
%\plotone{fig1.eps}
%\includegraphics[width=10cm]{fig1.eps}
\includegraphics{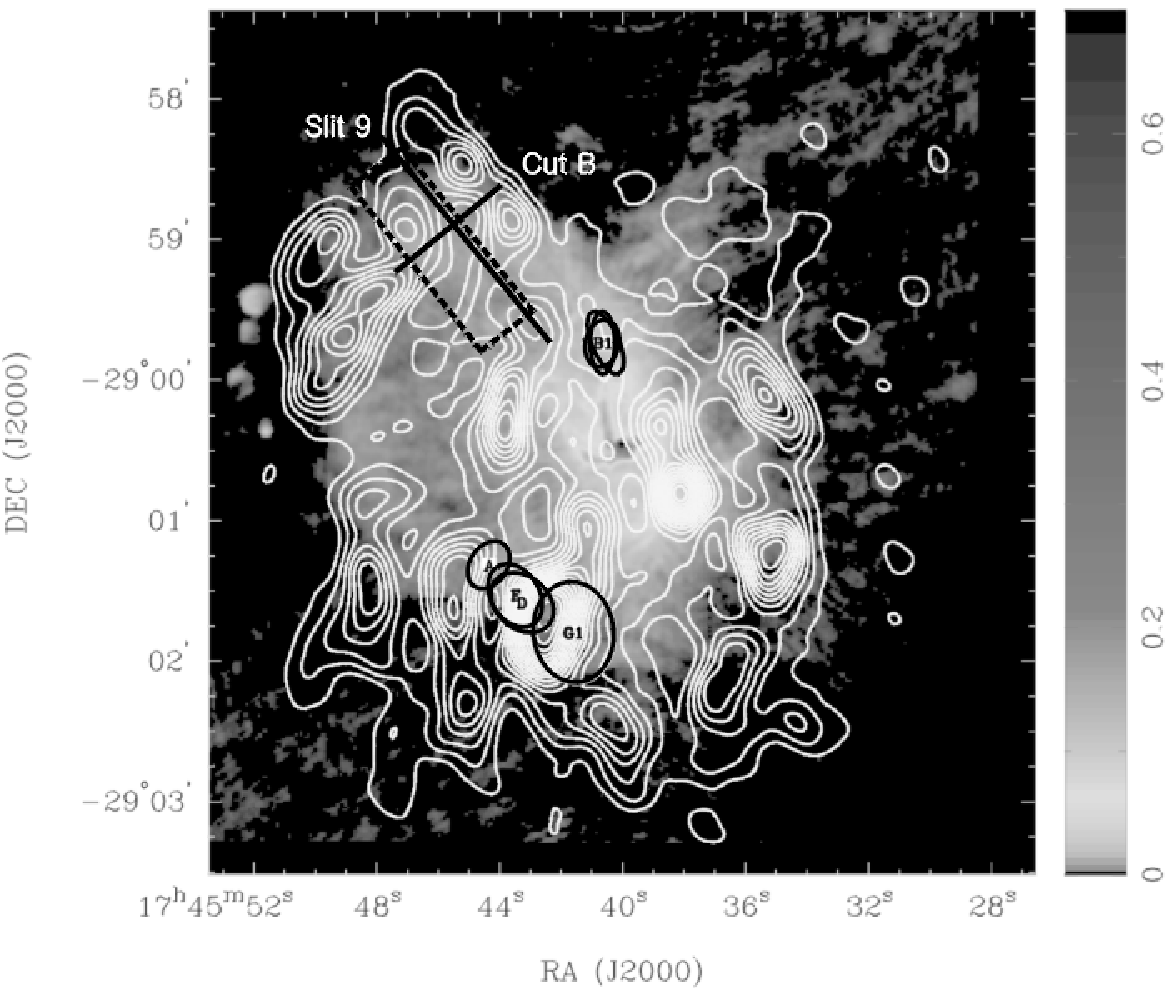}
\vspace{97mm}
\caption{Central $10 \times 15$ pc region of the Galaxy.
Contours representing the velocity-integrated map of NH$_3$(3,3) emission
are overlaid on a 6 cm continuum image of Sgr A complex from \citet{mcg01}.
The black dot at the centre of image is Sgr A* and the mini spiral is Sgr A West.
The CND is traced by the brighter part of continuum surrounding them
and Sgr A East is seen by the outer part extended to the boundary of NH$_3$ contours.
The GMC M-0.02-0.07 lies to the east of this region.
The dashed box at the northeastern edge of Sgr A East
encloses the $90 \times 27$ arcsec region observed in H$_2$.
The two solid lines in the box are the
locations from which the H$_2$ spectra shown in this
paper were extracted.  Cut B is made across the 10 slits while the NE--SW
cut, perpendicular to the edge of Sgr A East, is Slit 9.
Letters mark the positions of OH(1720 MHz) masers with error ellipses
scaled up by a factor of 15 \citep{yus99a}.
\label{fig_target}}
\end{figure*}

\citet{yus87} suggested that a different kind of explosive event could
create Sgr A East. The energy required to make a huge shell such as Sgr A
East has been associated with a `hypernova' \citep*{woo99}. \citet{kho96}
hypothesize that Sgr A East may be the remnant of a solar mass star
tidally disrupted by a massive black hole.  Their model can naturally
explain the elongated shape of Sgr A East as well as the extreme
energetics. However, from their observation with the {\it Chandra X-ray
Observatory}, \citet{mae02} suggest that Sgr A East should be classified
as a metal-rich `mixed morphology' supernova remnant .  They argue that
the model of \citet{kho96} cannot reproduce the metal-rich abundances
observed at the centre of Sgr A East. They also conclude that a single
Type II supernova explosion
with an energy of $10^{51}$ ergs into an homogeneous ambient medium with a
density of $10^{3} \; \cmv$ can most simply explain the results of both
radio and X-ray observations, and thus that the extreme energy of $\sim
10^{52}$ ergs is not required.

In principle, the energy of the explosive event can be directly measured
by studying regions where Sgr A East is colliding with ambient
interstellar material. By tracing the dynamics of molecular gas, an
interaction between the eastern part of Sgr A East and the giant molecular
cloud (GMC) M-0.02-0.07 (also known as the `$50~\kms$ cloud') has been
inferred \citep*{gen90,ho91,ser92,mez96,nov99,coi00}. Recent observations
of NH$_3$(3,3) emission in the region show that Sgr A
East impacts material to the north and west as well
(see Fig.~\ref{fig_target}) \citep*{mcg01}.  As
direct evidence of this interaction, several 1720 MHz OH masers, which are
a good diagnostic of the continuous, or C-type, shock excitation
\citep*{fra96,war99}, have been detected along the southern edge of Sgr A
East and to the north of the circum-nuclear disc (CND) \citep{yus96}.

\citet{war99} and \citet{yus99b,yus01} detected H$_2$ line emission in
regions where OH-masers have been detected. In most cases H$_2$ line
emission arises either from thermal excitation (e.g. by shock heating) or
from non-thermal excitation by far-UV absorption \citep{bla87,bur92,pak98}.
One can in principle distinguish between these two mechanisms by comparing
near-infrared (near-IR) line intensities.  The $\hto$ / $\hozs$ ratio has
been an effective discriminant in a number of shocked regions (where the
ratio should be $\le 0.3$) and photodissociation regions (PDRs)
(where it is about 0.5 -- 0.6).  However, a `thermal' line
ratio can be observed in a PDR -- even though fluorescence is the dominant
excitation mechanism -- if the gas density is high ($\ge 10^5 \; \cmv$;
\citealt{ste89}). \citet{gat84} observed the CND and concluded that the
H$_2$ molecules are excited by collisions, while the results for larger
regions (about $2 \times 2$ deg$^2$) by \citet*{pak96a,pak96b} are
consistent with non-thermal excitation. The interpretation of
\citet{war99} and \citet{yus99b,yus01} that the line emission in Sgr A East
is thermal is supported by the presence of the 1720 MHz OH masers.  It is
therefore likely that Sgr A East is indeed driving shocks into the
adjacent GMCs to the south and into the CND.

\begin{figure*}
%\plotone{fig2.eps}
%\includegraphics[]{fig2.eps}
\includegraphics{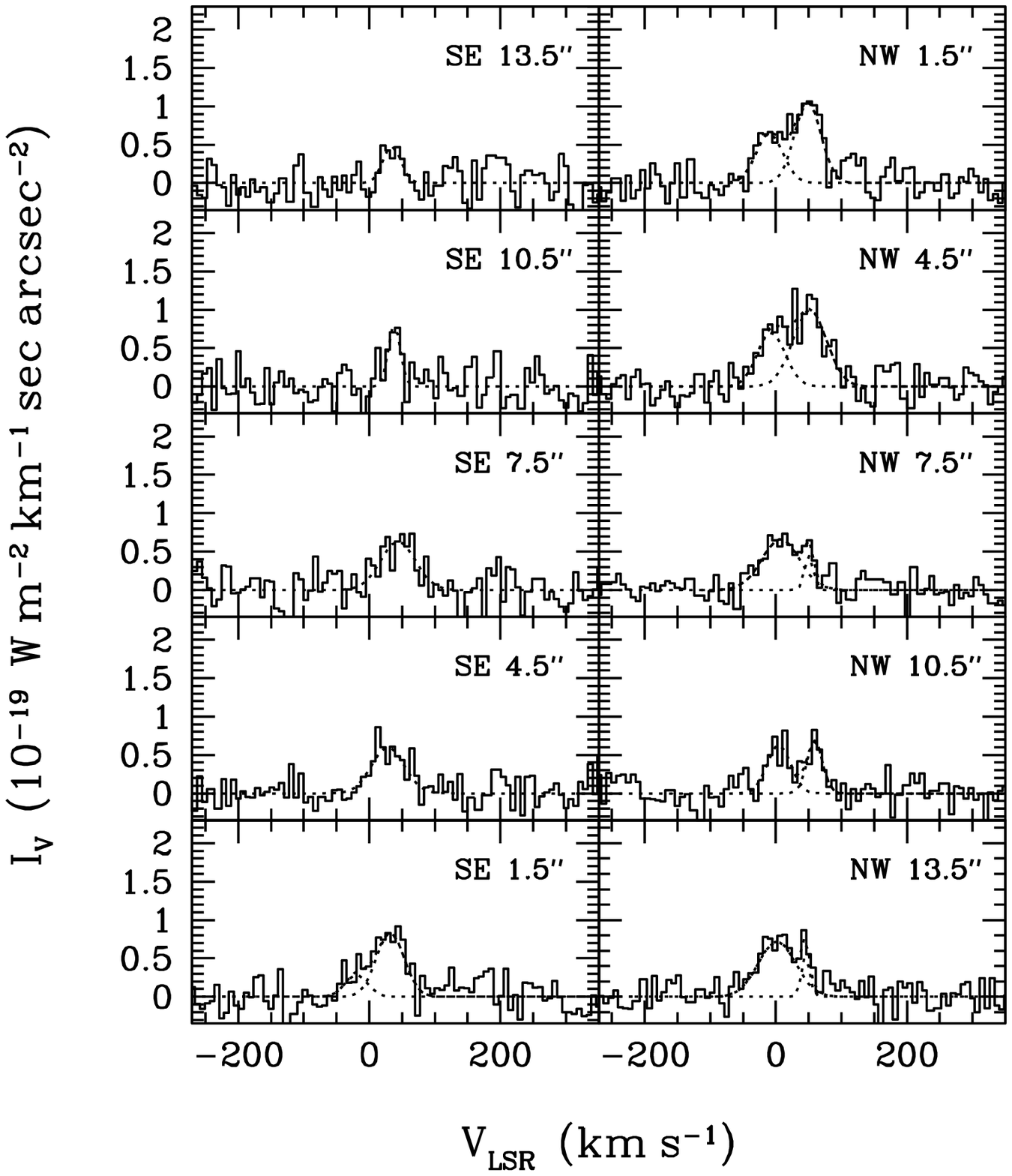}
\vspace{134mm}
\caption{$\hoz$ spectra at 10 positions along Cut B.  Indicated positions
are relative to the centre of the cut ($\rm \alpha = 17^h 45^m 46\fs1, ~
\delta = -28\degr59\arcmin01\arcsec$; J2000).  The dotted lines
are Gaussian fits to the observed line profiles.
The spectra are not corrected for instrumental broadening.
\label{fig_spect_bcut}}
\end{figure*}

The fields observed by \citet{war99} and \citet{yus99b,yus01} are
restricted to the vicinity of the CND and cover only some of the regions
where interaction of the Sgr A East shell with surrounding material is
expected. Before one can hope to estimate the energy released in the event
that created Sgr A East, it is necessary to observe additional interaction
regions in diagnostic lines of H$_2$ at high spectral resolution. In this
paper we present velocity-resolved, near-IR H$_2$ observations at the
northeastern boundary of Sgr A East.  By measuring $\hto$ / $\hozs$ line
ratios and line profiles simultaneously we aim to study the excitation and
kinematics of the interaction between Sgr A East and M-0.02-0.07.

\section{Observation and Data reduction}

We observed the $\hoz ~ (\lambda = 2.1218 \micron)$ and the $\hto ~
(\lambda = 2.2477 \micron)$ spectra at the 3.8 m United Kingdom Infra-Red
Telescope (UKIRT) in Hawaii on 2001 August 3 and 4 (UT), using the Cooled
Grating Spectrometer 4 (CGS4; \citealt{mou90}) with a 31 l/mm
echelle grating, 300 mm focal
length camera optics and a two-pixel-wide slit. The spatial resolution
along the
slit was 0.90 arcsec for $\hoz$ with the grating angle of 64\,\fdg691 and
0.84 arcsec for $\hto$ with 62\,\fdg127, respectively; the slit widths on
the sky were 0.83 and 0.89 arcsec, respectively, for these two
configurations. The slit length is $\sim$~90 arcsec. The instrumental
resolutions, measured from Gaussian fits to sky lines in our raw data,
were $\sim 17~\kms$ for $\hoz$ and $\sim 19~\kms$ for $\hto$, respectively.

Ten parallel slit positions were observed, sampling a $90 \times 27$
arcsec area on the northeastern boundary of Sgr A East. The slit was
oriented $40 \degr$ east of north for each measurement; adjacent slit
positions were separated by 3 arcsec perpendicular to the slit axis. The
coordinates at the centre of the observed area are $\rm \alpha = 17^h 45^m
45\fs9, ~ \delta = -28\degr59\arcmin05\arcsec$ (J2000) (see
Fig.~\ref{fig_target}).  The southwestern part of this region includes the
`outer H$_2$ clumps' from which H$_2$ emission was detected by
\citet{yus01}.  Only the ninth slit position, hereafter called `Slit 9',
was observed in both $\hoz$ and $\hto$.  The telescope was nodded between
object and blank sky positions every 25 minutes, to subtract the
background and telluric OH line emission. The sky positions were offset by
about 2\,\fdg5 ($\Delta \alpha = -2\,\fdg03, ~ \Delta \delta = 0\,\fdg85$)
from the on-source positions.

\begin{figure*}
%\plotone{fig3.eps}
%\includegraphics[]{fig3.eps}
\includegraphics{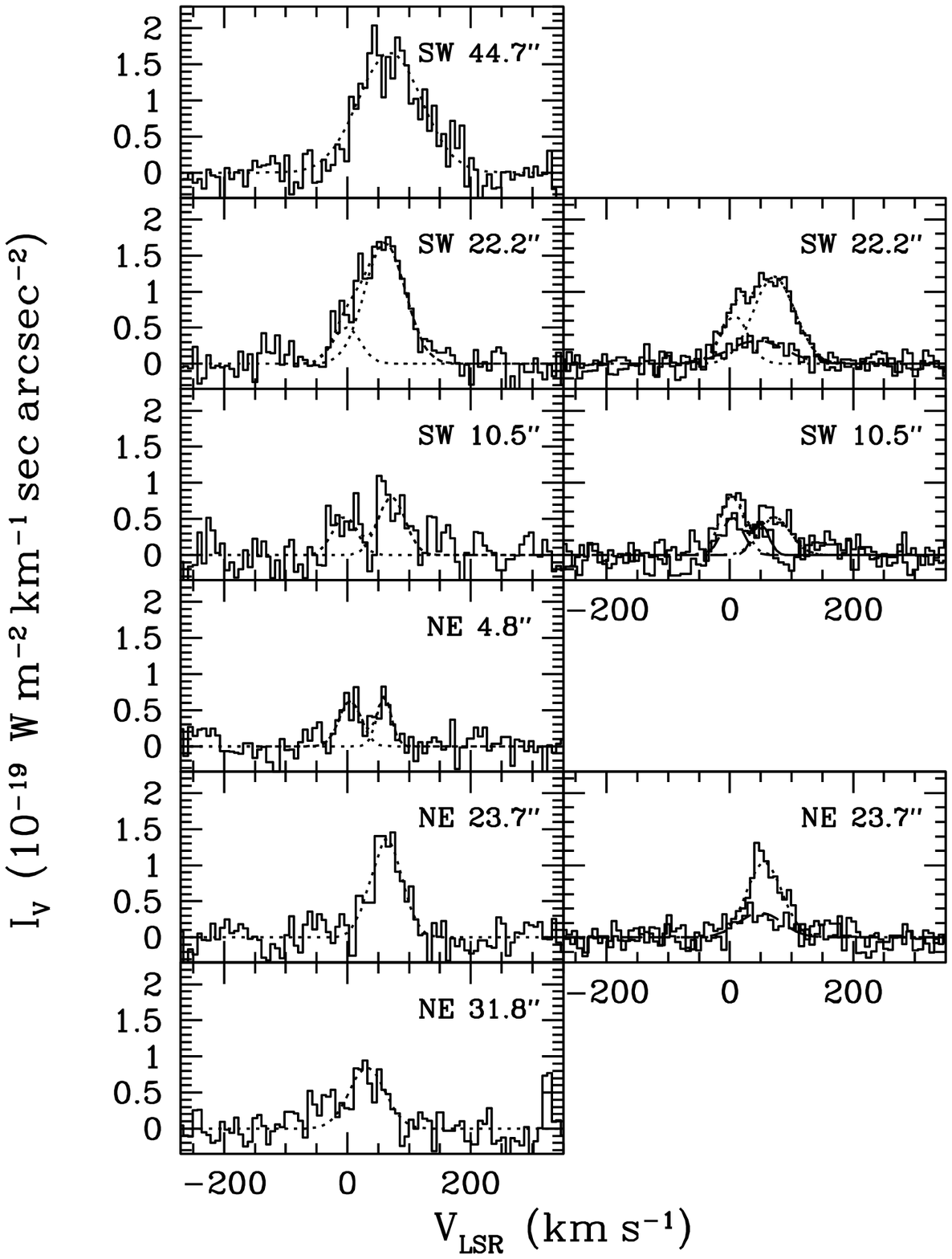}
\vspace{153mm}
\caption{$\hoz$ and $\hto$ spectra from six positions along Slit
9. Indicated positions are relative to $\rm \alpha
= 17^h 45^m 45\fs3, ~ \delta = -28\degr58\arcmin58\arcsec$; J2000.  Left
panels show the $\hoz$ spectra.  The right three panels present both the
$\hoz$ and $\hto$ spectra from the positions where $\hto$ emission is
detected; these are averaged over 3.4 arcsec on the sky to improve the S/N
ratios.  The other aspects are the same as Figure~\ref{fig_spect_bcut}.
\label{fig_spect_nw12}}
\end{figure*}

Initial data reduction steps, involving bias-subtraction and flat-fielding
(using an internal blackbody lamp), were accomplished by the automated
Observatory Reduction and Acquisition Control ({\sc orac}) pipeline at
UKIRT. {\sc iraf} was used for the remainder of the reduction.  We
corrected the spectral distortion along the dispersion axis using the
spectrum of the standard star HR 6496 as a template. The sky OH lines were
then used to correct for spatial distortion perpendicular to this axis and
also for wavelength calibration. We also corrected for the motions of the
Earth and Sun in order to determine local standard of rest (LSR)
velocities.

Only part of the flux from a standard star is detected due to the narrow
slit; hence for proper flux calibration the measured signal must be
corrected.  We assumed a circularly symmetric PSF for the star,
based on the flux profile along the slit length, to estimate the missing
flux.  The correction factor, which varies with the seeing, ranged from
2.06 to 2.56. Near-IR
emission from the Galactic centre is attenuated by interstellar
material in the foreground (mostly 4 -- 8 kpc from the Galactic centre)
and by material in the Galactic centre itself.  Since we believe that the
H$_2$ line emission originates from the surface of the cloud, we ignore
the latter \citep{pak96a, pak96b} and correct only for foreground
extinction; which we assume to be $A_K = 2.5$ mag \citep*{cat90}.

\begin{figure*}
%\plotone{fig4.eps}
%\includegraphics[]{fig4.eps}
\includegraphics{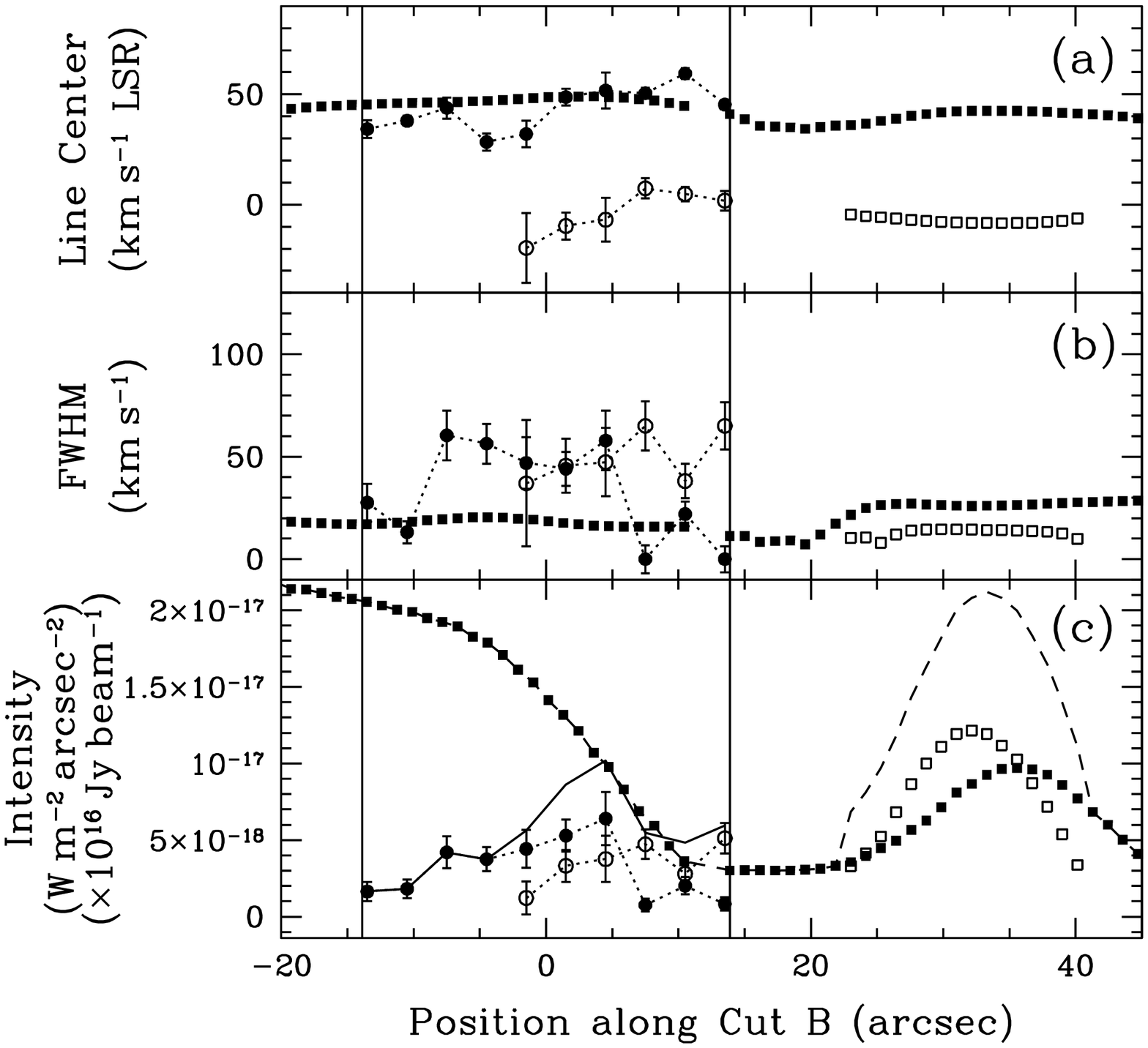}
\vspace{116mm}
\caption{Derived line parameters for the spectra along Cut B:
(a) line center velocity; (b) line width; and (c) integrated line
intensity.  Indicated positions are as in Fig.~\ref{fig_spect_bcut};
positive towards NW. The filled circles and
the open circles represent the +50 $\kms$ component and the 0 $\kms$
component of $\hoz$ spectra, respectively, and the filled and open squares
(the +50 $\kms$ and 0 $\kms$ component) of the NH$_3$(3,3), from
\citet{mcg01}.  In the panel (c), the solid and dashed lines denote the
total intensity of $\hoz$ and NH$_3$(3,3), respectively, the latter scaled
by $10^{-16}$.  The range of our H$_2$ observation is denoted by two
vertical lines.
\label{fig_param_bcut}}
\end{figure*}

\section{Results and Discussion}

\subsection{$\hoz$ emission} \label{emission}

Bright (1.6 -- 21 $\intensity$) H$_2$ emission was detected from most of
the observed region ($90 \times 27$ arcsec) along the northeastern
boundary of Sgr A East. We created a SE--NW cut parallel to the boundary
of Sgr A East (`Cut B') by extracting spectra from each of the ten slit
positions. A second cut perpendicular to the boundary (i.e., SW--NE) is
composed of six positions from Slit 9.
The positions of Cut B and Slit 9 are marked in Fig.~\ref{fig_target}.
Along Cut B, the intensity of NH$_3$ emission varies dramatically;
we can investigate both high and low density regions along this cut.
Slit 9 is the only slit observed both in $\hoz$ and $\htos$,
the line ratio of which we can use to constrain models for the excitation of H$_2$.
In this paper, we present the results of these two cuts, rather than the whole data set,
as a preliminary report. We aim to concentrate on the excitation mechanism
of the detected H$_2$ emission, and the map the structure and kinematics
perpendicular to and parallel to the boundary.
From the sixteen $\hoz$ spectra in
Figs.~\ref{fig_spect_bcut} \& \ref{fig_spect_nw12}
we measure line centres and line widths along the interaction region.
Each spectrum is well fitted by one or two Gaussian components.

Figs.~\ref{fig_param_bcut} \& \ref{fig_param_nw12} show the
distributions of the derived $\hoz$ line parameters along Cut B and Slit
9, respectively.  For direct comparison, we include in these figures data
from the NH$_3$(3,3) observations of \citet{mcg01}; the NH$_3$
emission essentially traces the cool ($\la 100$ K), dense ($10^5~\cmv$),
cloud material.  From these data we note the following.

\begin{itemize}

\item{In Figs.~\ref{fig_param_bcut}(a) \& \ref{fig_param_nw12}(a) there
are two velocity components, at $V_{\rm LSR} \sim 0~\kms$ and $\sim +50
~\kms$. Both components are evident in H$_2$ and NH$_3$,
although the 0 $\kms$ features are not spatially coincident. The variation in
the velocity of either component is less than $\sim 20~\kms$ along both
Slit 9 and Cut B.}

\item{The H$_2$ line widths in Figs.~\ref{fig_param_bcut}(b) \&
\ref{fig_param_nw12}(b) are much broader than the NH$_3$ widths and show
no obvious trend along either Slit 9 or Cut B.}

\item{Along Cut B, the distribution of the total intensity of H$_2$
(the solid line in Fig.~\ref{fig_param_bcut}(c))
is quite different from the NH$_3$ (the
dashed line; the units are arbitrarily scaled).
The NH$_3$ emission increases to the southeast as the cut passes deeper
into the body of the GMC M-0.02-0.07, but the H$_2$ emission decreases in
this direction.
The decrease in H$_2$ may be explained either by
exhaustion of the source of excitation (e.g. shock energy or UV photons)
or by obscuration, at the inner, more dense, regions of the cloud,
or by the geometry of the interaction region (see Section 3.3).}

\begin{figure*}
%\plotone{fig5.eps}
%\includegraphics[]{fig5.eps}
\includegraphics{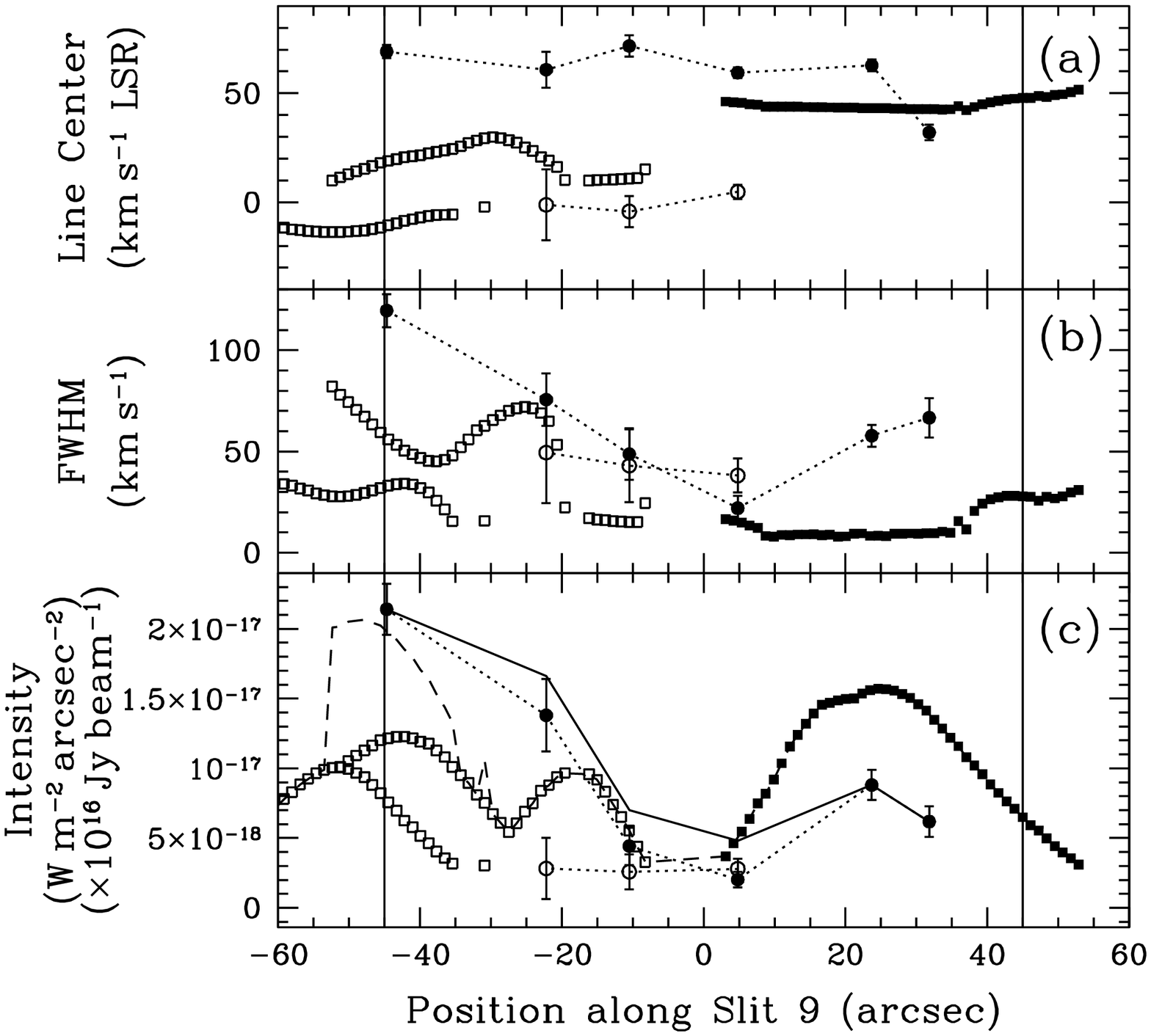}
\vspace{116mm}
\caption{Derived line parameters for the spectra along Slit 9.
Indicated positions are as in Fig.~\ref{fig_spect_nw12};
positive towards NE. Identities of plotted quantities are
as in Fig.~\ref{fig_param_bcut}.
The decrease in NH$_3$ flux at positions greater than $+30\arcsec$
is a result of reduced sensitivity at the edge of the mosaic \citep{mcg01}.
\label{fig_param_nw12}}
\end{figure*}

\item{Along Slit 9, the distribution of the H$_2$ intensity (solid line) in
Fig.~\ref{fig_param_nw12}(c) is generally similar to that of NH$_3$
(dashed line).
It should be noted that
NH$_3$ is attenuated at positions greater than $+30\arcsec$
by the edge of the primary beam in the VLA mosaic \citep{mcg01}.
There is a small discrepancy between NH$_3$ and H$_2$
between $+10\arcsec$ and $+30\arcsec$
which may be the result of the same effects discussed in Cut B.
Note that Slit 9 covers a very large spatial scale (90 arcsec).
The brightest emission along Slit 9, between offsets $-20\arcsec$ to
$-45\arcsec$ (the southwestern part), arises from
the outer H$_2$ clumps of \citet{yus01}.}

\end{itemize}

\subsection{H$_2$ excitation} \label{excite}

The $\hto$ line was detected at three locations along
Slit 9, at positions NE $23\farcs7$, SW $10\farcs5$, and SW $22\farcs2$
relative to the centre of the slit ($\rm \alpha = 17^h 45^m 45\fs3, ~
\delta = -28\degr58\arcmin58\arcsec$; J2000).  From these data we measured
line ratios ($\hto$ / $\hozs$) of $0.40\pm0.12$, $0.51\pm0.17$, and $0.27\pm0.07$, respectively
(see Fig.~\ref{fig_spect_nw12}). At other positions only the $\hoz$ line was
detected, with 3$\sigma$ upper limits to the ratio of 0.5, 0.6, and 0.1 at
offsets of NE $31\farcs8$, NE $4\farcs8$, and SW $44\farcs7$ along Slit 9,
respectively.

Fluorescent excitation in a
low-density PDR ($\rm n(H_2) < 5 \times 10^4 \; \cmv$) should yield a
ratio of about 0.6.  A lower ratio is expected in a more dense PDR
environment \citep{bla87}, or in a shock. There are two basic types of
shock; `jump' or J-type and `continuous' or
C-type (see \citealt{dra93} for a review). A J-type shock is formed in a
highly ionized or weakly magnetized gas. Fluid parameters such as density
and temperature undergo a discontinuous change (jump) at the shock front
where the molecules may be dissociated.  J-type shocks (with
velocities greater than about 24 $\kms$) will completely dissociate the
molecules \citep{kwa77}; H$_2$
emission occurs from a warm, recombination plateau in the post-shock
region.  J-type shocks typically produce low line intensities and
$\hto ~ / ~ \hozs$ line ratios as large as 0.5 are possible
\citep{hol89}.  At lower shock velocities, below the H$_2$ dissociation
speed limit, J-type shocks may yield much lower line ratios; $<0.3$
\citep{smi95}.  In a C-type shock, where the magnetic field softens the
shock front via ion-magnetosonic wave propagation so that the fluid
parameters change continuously across the shock front, the H$_2$
dissociation speed limit is much higher ($\sim$ 45 $\kms$; depending on
the density and magnetic field strength in the pre-shock gas).
Smaller line ratios of about 0.2 are then predicted \citep{smi95,kau96}.

From the observed ratios alone we are not able to unambiguously
distinguish between excitation mechanisms.  Our results can either be
explained by fast J-type shocks or dense PDRs, or by a combination of
fluorescence and either C-type shocks or slow J-type shocks, since the
higher line ratios associated with fluorescence will be tempered by
the low $\hto$ intensities associated with collisional excitation
in shocks.

To help distinguish between the H$_2$ excitation mechanisms, we consider
kinematic information and the spatial variation of the line ratio along
slit 9 . At most positions in Figs.~\ref{fig_param_bcut}(b) \&
\ref{fig_param_nw12}(b), the H$_2$ line widths are high (typically 40 --
70 $\kms$, but as high as 120 $\kms$ in some positions). This suggests
shock excitation and turbulent motions in the gas and tends to exclude the
pure fluorescence models (in which the H$_2$ line emission generally
arises from the stationary gas at the edges of neutral clouds illuminated
by Far-UV photons from early-type stars).  However, in other shocked
regions the line ratio is found to be constant over a wide range of $\hoz$
intensities and spatial positions \citep*{dav95,ric95}, although this is
not necessarily predicted from theory \citep{dra93}. Conversely, in a PDR
the ratio is sensitive to the incident FUV flux and the molecular gas
density (the $\hoz$ intensity increases but the $\hto$ / $\hozs$
ratio decreases with increasing gas
density or UV intensity; \citealt{usu96,tak00}).  Thus an unchanging
$\hto ~ / ~ \hozs$ ratio is
found in shocks, while a varying ratio is expected in the
pure fluorescent case.  The measured line ratio and the $\hoz$ intensity
in Fig.~\ref{fig_param_nw12}(c), show evidence of an anti-correlation in
our data, as expected in dense PDRs.  Although the wide line profiles
point to shock excitation, fluorescence appears to play a significant role
at at least some locations.

Considering the kinematics further, we note that J-type shocks produce
narrow lines that peak at the velocity of the shock, while C-type shocks
produce broader lines which peak at the velocity of the pre-shock gas and
extend up to the shock velocity.  Figs.~\ref{fig_param_bcut}(a) \&
\ref{fig_param_nw12}(a) show that there are two velocity components that
are similar in H$_2$ and NH$_3$. The H$_2$ emission traces hot ($\sim
2000$ K) gas and the NH$_3$ cool ($\la~100$ K) gas. Thus, if we assume that
shocks are driven by Sgr A East into cold molecular gas, whose velocities
are given by the NH$_3$ data, then fast J-type shocks are inconsistent
with our results, due to the low peak velocities of the H$_2$ lines
relative to the molecular clouds.

In summary, then, the wide line profiles and low peak velocities indicate
C-type shock excitation.  However, the high values of the line ratio at
some positions along Slit 9 and the spatial variation in that ratio, point
to a fluorescent component to the excitation in some locations.  A
combination of C-type shocks and fluorescence (see e.g. \citealt*{fer97}) is
therefore the most reasonable explanation for the H$_2$ excitation.
For the fluorescence, the source of the UV radiation could be either
nearby early type stars or J-type shocks.
However, as noted above, we see no evidence of J-type shocks in our data.
Also, we cannot establish whether nearby stars are the source of the UV flux
due to the lack of information on where or how many early type stars
there are in the region.

\begin{figure}
\includegraphics{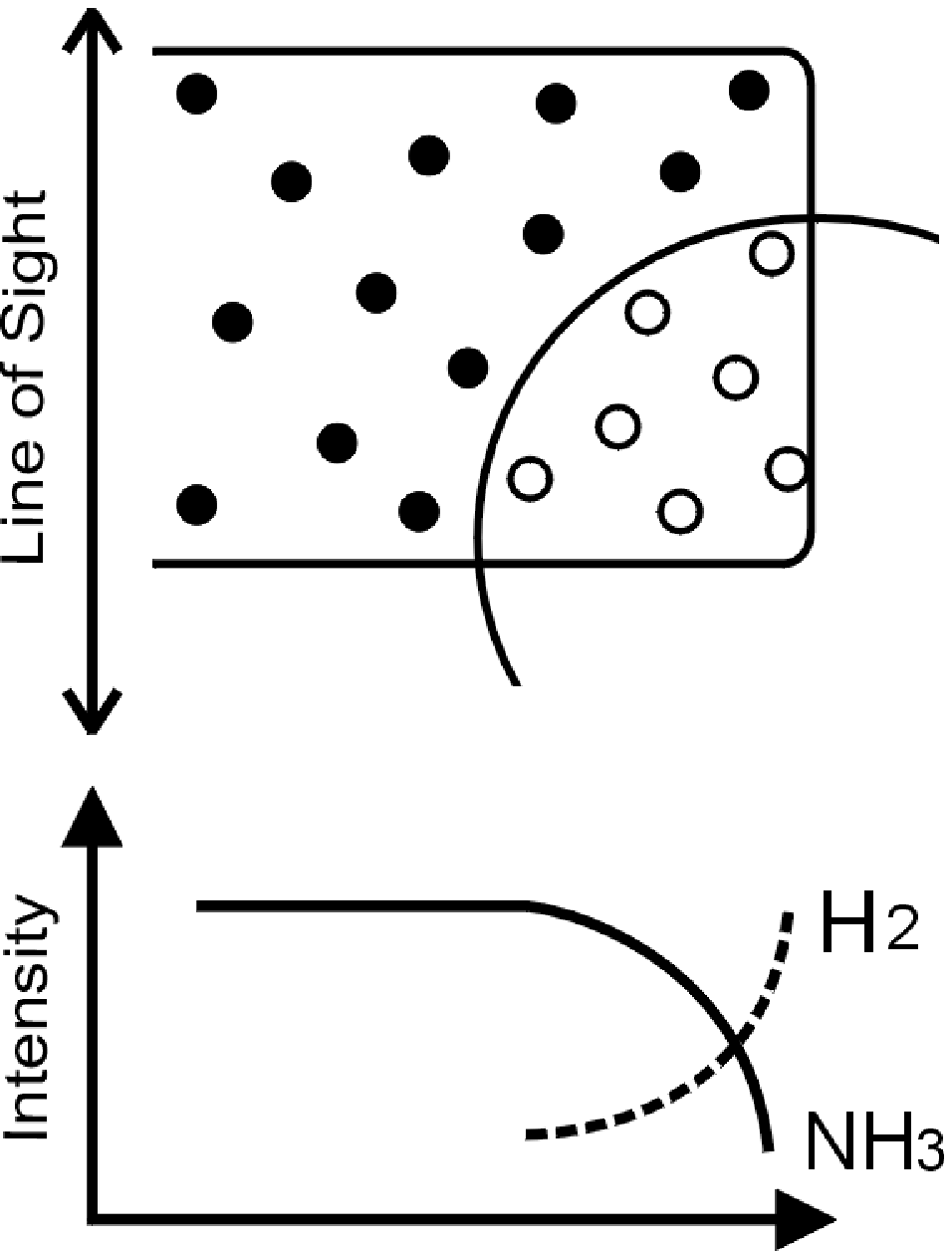}
\vspace{89mm}
\caption{Schematic diagram of Sgr A East and the GMC M-0.02-0.07
along the line of sight.
The large partial circle shows the boundary of Sgr A East
and the small circles enveloped in a square (M-0.02-0.07)
represent the clumpy structure of cloud (see Section 3.3).
Open circles are clumps that are being penetrated by shocks,
which have destroyed NH$_3$ but have excited the H$_2$ into emission.
The resulting intensity distribution shows similar trend with
the observed results (see the +50 $\kms$ components in
Figs.~\ref{fig_param_bcut}(c) \& \ref{fig_param_nw12}(c)).
\label{fig_structure}}
\end{figure}

\subsection{Structure of the interaction region} \label{structure}

The extended +50 $\kms$ component of the NH$_3$ emission traces the GMC
M-0.02-0.07, while the 0 $\kms$ component corresponds to the `northern
ridge' of \citet{mcg01} (see the NH$_3$(3,3) channel maps in their Fig.~4).
The NH$_3$ component at $\sim$ +20 $\kms$ seen in
Fig.~\ref{fig_param_nw12}(a) at negative offsets seems to trace hotter gas
according to the NH$_3$(2,2) to (1,1) line ratio map by \citet{mcg01}, which
they suggest may be the result of an impact by Sgr A East.

The difference in line width between NH$_3$ and H$_2$ as well as
the different spatial locations of their 0 $\kms$ features (Fig.~\ref{fig_param_nw12}(a))
indicate that the NH$_3$ and H$_2$ trace fundamentally
different components of the gas.
The narrow NH$_3$ line widths arise in ambient clouds.
The broader H$_2$ line emission traces shocked gas
where the NH$_3$ molecules are likely destroyed.
One can envision a situation in which Sgr A East is located adjacent to
both M-0.02-0.07 (the +50 $\kms$ component)
and the northern ridge (the 0 $\kms$ component),
with Sgr A East driving C-type shocks into these clouds.

On the other hand, the fact that the two H$_2$ velocity components overlap
along the line-of-sight, yet are roughly equally bright seems somewhat
unlikely, as one must be attenuated by the molecular cloud associated with
the other. The rough equality might be explained by clumpiness of
the foreground cloud, regardless of which one is in the foreground,
where a small filling factor of high density clumps are embedded within
a less dense medium \citep*{bur90}.
The size of such clumps seems to be $10^{-4}$ -- $10^{-3}$ pc \citep*{gar87,chu87},
which is smaller than our resolution of $\sim$ 1 arcsec
($\sim 4\times10^{-2}$ pc at the distance of $\sim$ 8.5 kpc to the Galactic center).
As illustrated in Fig.~\ref{fig_structure}, this clumpy structure could explain both the
observed decrease of NH$_3$ line emission and the observed increase of H$_2$
line emission toward the edge of the cloud
(see the +50 $\kms$ components in
Figs.~\ref{fig_param_bcut}(c) \& \ref{fig_param_nw12}(c)).

%\citet{coi00} suggested that Sgr A East is located in front of M-0.02-0.07
%from their NH$_3$ observation. However, some more recent observational results
%argue that M-0.02-0.07 be in front of Sgr A East \citep*{kar03,vol03}.
%Our Near-IR data cannot distinguish the positional relationship
%between Sgr A East and M-0.02-0.07 assuming the clumpy structure of the GMC.
%Because the cloud is optically thin in Near-IR
%though the small clumps in it are optically thick by the dust particles surviving from shocks,
%the H$_2$ emission from the rear part of cloud can be observed
%even if the GMC is located in front of Sgr A East.

\section{Conclusion}

We observed the northeastern part of the Sgr A East shell in order to
investigate its interaction with the GMC M-0.02-0.07. The bright $\hoz$
emission is strong evidence that Sgr A East is physically adjacent
to, and interacting with, M-0.02-0.07.

By comparing the relative intensities of $\hoz$ and $\hto$ emission, the
distribution of the $\hto$ / $\hozs$ line ratio, and the radial velocities
of the H$_2$ emission, we can to some extent distinguish between
excitation mechanisms for the H$_2$.  The line ratios tend to support emission in
either fast J-type shocks or a dense PDR.  However, on considering the
bright $\hoz$ intensity, the large line widths, and the spatial variation
in the line ratio, we conclude that a combination of C-type shocks and
fluorescence is required. The presence of shocks is direct evidence that
Sgr A East is driving into the surrounding material, and is consistent
with the detection of 1720 MHz OH masers to the north of the CND and to
the south of Sgr A East \citep{yus96}.
Very recently \citet{kar03} detected the 1720 MHz OH masers also at
two positions near our target region, which is more direct evidence
supporting our conclusion on the C-type shocks.

The H$_2$ emission covers most parts of our targeted region ($90 \times
27$ arcsec).  The line profiles are made up of two velocity components
both of which extend over a significant portion of the region ($15 \times
27$ arcsec). We find that the NH$_3$(3,3) emission lines observed by
\citet{mcg01} also show a similar kinematic structure, with almost the same
velocities. We suggest that the H$_2$ line emission arises at the
interfaces between Sgr A East and two independent molecular clouds, with
line-of-sight velocities of $\sim$ +50 $\kms$ (M-0.02-0.07) and
$\sim$ 0 $\kms$ (the northern ridge).
Both the observed two velocity components of the H$_2$ emission
and the difference in the intensity distributions between the H$_2$ and NH$_3$ emission
can be understood if the molecular clouds are composed of small dense
clumps with a very small filling factor.

To study the origin and evolution of Sgr A East, it would be important
to know the total H$_2$ luminosity and the total cooling rate (based on that)
over the interaction region, which could be compared with those of well studied SNRs.
However, it is very difficult
to estimate them with the small amount of information we have at present.
Given the uncertainty in the emission mechanisms, even estimating the total H$_2$
luminosity in the small mapped region would be difficult,
without considering the entire interaction region.
It would be premature for us to estimate the required energy to make the Sgr A East shell.
We will be able to do this in the future,
after observations of more of the interaction region.

\section*{Acknowledgments}

We give special thanks to Young-Sam Yu and Tae-Hyun Kim for their help
with the observation.  Fig.~\ref{fig_target} is reproduced from Fig.~10 of
\citet{mcg01} by permission of the AAS.  The United Kingdom Infrared
Telescope is operated by the Joint Astronomy Centre on behalf of the U.K.
Particle Physics and Astronomy Council. This work was financially
supported by the BK21 Project of the Korean Government. TRG's research is
supported by the Gemini Observatory, which is operated by the Association
of Universities for Research in Astronomy, Inc., on behalf of the
international Gemini partnership of Argentina, Australia, Brazil, Canada,
Chile, the United Kingdom and the United States of America.

\bsp

\label{lastpage}

\end{document}